\documentclass[10pt,a4paper]{article}
\usepackage{amsmath,amssymb,amsfonts}
\usepackage{graphicx}
\usepackage{hyperref}
\usepackage[margin=0.75in]{geometry}
\usepackage{float}
\usepackage{bm}
\usepackage{multirow}

\usepackage{setspace}
\usepackage{titlesec}
\titlespacing*{\section}{0pt}{8pt}{4pt}
\titlespacing*{\subsection}{0pt}{6pt}{3pt}

\AtBeginDocument{
  \setlength{\abovedisplayskip}{4pt}
  \setlength{\belowdisplayskip}{4pt}
  \setlength{\abovedisplayshortskip}{2pt}
  \setlength{\belowdisplayshortskip}{2pt}
}

\let\oldbibliography\thebibliography
\renewcommand{\thebibliography}[1]{%
  \oldbibliography{#1}%
  \setlength{\itemsep}{0pt}%
  \setlength{\parskip}{0pt}%
}

\title{Preliminary sonification of ENSO using traditional Javanese gamelan scales}

\author{Sandy H. S. Herho$^{1,2,3,4}$, Rusmawan Suwarman$^{5,*}$, Nurjanna J. Trilaksono$^{5}$, \\Iwan P. Anwar$^{6}$, and Faiz R. Fajary$^{5}$}

\date{}

\begin{document}
\maketitle

\begin{center}
\small
$^{1}$Department of Earth and Planetary Sciences, University of California, Riverside, CA 92521, USA\\
$^{2}$School of Systems Science and Industrial Engineering, State University of New York (SUNY), Binghamton, NY 13902, USA\\
$^{3}$Ronin Institute for Independent Scholarship 2.0, Sacramento, CA 95816, USA\\
$^{4}$Center for Agrarian Studies, Bandung Institute of Technology (ITB), Bandung, West Java 40132, Indonesia\\
$^{5}$Atmospheric Science Research Group, Bandung Institute of Technology (ITB), Bandung, West Java 40132, Indonesia\\
$^{6}$Applied and Environmental Oceanography Research Group, Bandung Institute of Technology (ITB), Bandung, West Java 40132, Indonesia\\
$^{*}$e-mail: rusmawan@itb.ac.id
\end{center}

\begin{abstract}
\noindent Sonification---the mapping of data to non-speech audio---offers an underexplored channel for representing complex dynamical systems. We treat El Ni\~{n}o-Southern Oscillation (ENSO), a canonical example of low-dimensional climate chaos, as a test case for culturally-situated sonification evaluated through complex systems diagnostics. Using parameter-mapping sonification of the Ni\~{n}o 3.4 sea surface temperature anomaly index (1870--2024), we encode ENSO variability into two traditional Javanese gamelan pentatonic systems (\textit{pelog} and \textit{slendro}) across four composition strategies, then analyze the resulting audio as trajectories in a two-dimensional acoustic phase space. Recurrence-based diagnostics, convex hull geometry, and coupling analysis reveal that the sonification pipeline preserves key dynamical signatures: alternating modes produce the highest trajectory recurrence rates, echoing ENSO's quasi-periodicity; layered polyphonic modes explore the broadest phase space regions; and the two scale families induce qualitatively distinct coupling regimes between spectral brightness and energy---predominantly anti-phase in \textit{pelog} but near-independent in \textit{slendro}. Phase space trajectory analysis provides a rigorous geometric framework for comparing sonification designs within a complex systems context. Perceptual validation remains necessary; we contribute the dynamical systems methodology for evaluating such mappings.
\end{abstract}

\noindent\textbf{Keywords:} sonification, ENSO, gamelan, phase space analysis, recurrence quantification, complex systems

\section{Introduction}

The El Ni\~{n}o-Southern Oscillation (ENSO) is among the most thoroughly studied complex dynamical systems in geophysics. It exhibits hallmark features of low-dimensional nonlinear dynamics: irregular oscillations on a 2--7 year timescale, amplitude asymmetry between warm and cold phases, phase-locking to the annual cycle, and sensitivity to stochastic forcing \cite{mcphaden2006enso,trenberth1997definition,strogatz2015nonlinear}. These properties emerge from coupled ocean-atmosphere feedbacks---the Bjerknes mechanism, delayed oscillator dynamics, and recharge-discharge processes---that generate quasi-periodic behavior from deterministic nonlinear interactions modulated by noise \cite{an2005nonlinearity}. ENSO is, in short, a paradigmatic example of a complex system whose temporal evolution invites analysis through the tools of dynamical systems theory \cite{kantz2003nonlinear}.

Representing such systems for diverse audiences poses a fundamental challenge. Traditional visualization presupposes graphical literacy and engages a single sensory channel \cite{Sawe2020}. Sonification---the systematic transformation of data into non-speech audio---offers a complementary modality by leveraging the human auditory system's sensitivity to temporal structure, pattern recurrence, and multi-stream processing \cite{hermann2011sonification,Kramer1999}. Critically, sonification itself constitutes a complex mapping: a nonlinear transformation between the state space of the source system and an acoustic parameter space, where the dynamical properties of the output depend jointly on the input signal, the mapping function, and the musical framework employed. Whether and how well sonification preserves the dynamical structure of the source system is a question amenable to the tools of complex systems science.

Yet climate sonification has been pursued almost exclusively within Western musical frameworks---equal temperament, common-practice harmony, orchestral timbres---without examining how alternative tonal systems might encode dynamical structure differently \cite{GarciaBenito2025,Worrall2019}. This is a missed opportunity. Music itself exhibits complex systems characteristics: $1/f$ spectral scaling in pitch and loudness fluctuations, long-range correlations, hierarchical temporal organization, and emergent structure from simple generative rules \cite{voss1975,levitin2012musical}. Different musical traditions realize these properties through different structural mechanisms. Javanese gamelan, in particular, organizes time through nested cyclical structures rather than linear progression \cite{Becker1981}, uses stratified polyphony where instruments operate at different temporal densities \cite{Perlman2004}, and employs two principal pentatonic tuning systems (\textit{pelog} and \textit{slendro}) with distinctive interval distributions \cite{surjodiningrat1972tone,tenzer2000gamelan}. These features---cyclicity, multi-scale hierarchy, and non-uniform interval structure---map naturally onto the dynamical properties of ENSO.

Indonesia's geographic position at the nexus of Pacific and Indian Ocean interactions makes it among the most ENSO-sensitive regions on Earth \cite{mcphaden2006enso}. Strong El Ni\~{n}o events produce severe drought and forest fires; La Ni\~{n}a episodes bring catastrophic flooding, particularly across Java \cite{mcphaden2006enso}. The Javanese gamelan tradition, developed over more than a millennium on this same island, thus provides both geographic rationale and structural resonance for encoding ENSO dynamics.

In this study, we map ENSO variability through gamelan tuning systems into audio and ask what dynamical signatures survive the transformation. We construct phase space representations of the sonified output and apply recurrence-based diagnostics from nonlinear time series analysis \cite{zbilut1992embeddings,marwan2007recurrence} to characterize the resulting acoustic trajectories. The work shows that the two Javanese scales produce sonification variants with quantifiably different phase space geometries, and that trajectory analysis in normalized brightness-energy coordinates can systematically distinguish sonification designs from a dynamical systems perspective. We have not conducted perceptual experiments and make no claims about communicative effectiveness---the contribution here is methodological.

\section{Data}

We use the Ni\~{n}o 3.4 sea surface temperature (SST) anomaly index from the HadISST 1.1 dataset \cite{rayner2003global}, representing monthly mean SST anomalies averaged over the equatorial Pacific (5$^{\circ}$N--5$^{\circ}$S, 170$^{\circ}$W--120$^{\circ}$W). This index serves as the primary ENSO monitoring indicator \cite{trenberth1997definition}. The monthly time series spans January 1870 to December 2024 ($N = 1{,}860$ observations over 155 years), obtained from the NOAA Physical Sciences Laboratory \cite{noaa2024nino34}.

\section{Methods}

We first characterize the Ni\~{n}o 3.4 time series as a dynamical signal. Central tendency, dispersion, and distribution shape were quantified through standard moments. The skewness coefficient
\begin{equation}
\gamma_1 = \frac{1}{N} \sum_{i=1}^{N} \left(\frac{\theta_i - \bar{\theta}}{\sigma}\right)^3
\label{eq:skewness}
\end{equation}
captures the warm-phase amplitude asymmetry characteristic of ENSO's nonlinear dynamics \cite{an2005nonlinearity}, while the excess kurtosis
\begin{equation}
\gamma_2 = \frac{1}{N} \sum_{i=1}^{N} \left(\frac{\theta_i - \bar{\theta}}{\sigma}\right)^4 - 3
\label{eq:kurtosis}
\end{equation}
quantifies heavy-tailed behavior relative to Gaussian processes \cite{joanes1998comparing}. Here $\theta_i$ is the SST anomaly at month $i$, $\bar{\theta}$ the mean, $\sigma$ the standard deviation, and $N$ the total observations.

Temporal persistence---a defining feature of ENSO's ocean-thermal inertia---was assessed through the lag-1 autocorrelation
\begin{equation}
\rho_1 = \frac{\sum_{i=1}^{N-1}(\theta_i - \bar{\theta})(\theta_{i+1} - \bar{\theta})}{\sum_{i=1}^{N}(\theta_i - \bar{\theta})^2}.
\label{eq:autocorr}
\end{equation}
ENSO episodes were identified using operational criteria \cite{trenberth1997definition}: periods of at least three consecutive months with anomalies $\geq +0.5$\textdegree C (El Ni\~{n}o) or $\leq -0.5$\textdegree C (La Ni\~{n}a). Rate-of-change statistics from first-order differences $\Delta\theta_i = \theta_{i+1} - \theta_i$ characterize the typical and extreme rates of ENSO state evolution. All computations used NumPy \cite{harris2020array} and SciPy \cite{virtanen2020scipy}.

We frame sonification as a structured mapping $\mathcal{S}: \mathcal{X} \to \mathcal{Y}$ from the ENSO state space $\mathcal{X}$ to an acoustic parameter space $\mathcal{Y}$. The central question is what properties of the source dynamics survive the transformation. We implemented parameter-mapping sonification \cite{hermann2011sonification,walker2005mappings} using two Javanese pentatonic scales. The \textit{pelog} scale uses intervals $\Phi_{\text{pelog}} = \{0, 1, 3, 7, 8\}$ semitones (approximating C--D$\flat$--E$\flat$--G--A$\flat$), while the \textit{slendro} scale uses $\Phi_{\text{slendro}} = \{0, 2, 3, 7, 9\}$ (C--D--E$\flat$--G--A) \cite{tenzer2000gamelan,surjodiningrat1972tone}. Both have cardinality $|\Phi| = 5$, but their interval distributions differ: \textit{pelog} features unequal spacing (1, 2, 4, 1 semitones between successive degrees) while \textit{slendro} is more uniform (2, 1, 4, 2). This structural difference constitutes the primary independent variable in our design.

The mapping $\mathcal{M}: \mathbb{R} \to \{0,\ldots,127\}$ transforms anomalies $\theta(t) \in [-3, 3]$\textdegree C to MIDI note numbers through normalization and quantization:
\begin{equation}
\xi(t) = \frac{\theta(t) + 3}{6} \in [0, 1],
\label{eq:normalize}
\end{equation}
\begin{equation}
\psi(t) = \xi(t) \cdot N_{\text{steps}}, \quad N_{\text{steps}} = |\Phi| \cdot n_{\text{oct}} - 1 = 9,
\label{eq:continuous_position}
\end{equation}
where $n_{\text{oct}} = 2$ octaves yields 10 discrete pitch levels. Quantization employs rounding \cite{oppenheim1999discrete}:
\begin{equation}
s(t) = \lfloor \psi(t) + 0.5 \rfloor \in \{0, 1, \ldots, 9\}.
\label{eq:quantize}
\end{equation}
Decomposition into octave offset and scale degree via modular arithmetic gives
\begin{equation}
o(t) = \left\lfloor s(t) / |\Phi| \right\rfloor, \quad d(t) = s(t) \bmod |\Phi|,
\label{eq:decompose}
\end{equation}
and the MIDI note number is
\begin{equation}
\nu(t) = 60 + 12 \cdot o(t) + \Phi_{d(t)},
\label{eq:midi_note}
\end{equation}
where 60 corresponds to middle C. This establishes a monotonic map: warm anomalies produce higher pitches, preserving ordinal structure.

To encode both state magnitude and evolution rate---effectively projecting phase space information into a single audio parameter---MIDI velocity is computed as
\begin{equation}
v(t) = \text{clip}\!\left(80 + 20|\theta(t)| + 30\left|\Delta\theta(t)\right|,\; 40,\; 127\right),
\label{eq:velocity}
\end{equation}
where $\text{clip}(x,a,b) = \max(a, \min(x,b))$ and $\Delta\theta(t) = \theta(t) - \theta(t-1)$. The weighting of the rate term ($\beta_2 = 30$) above the magnitude term ($\beta_1 = 20$) makes rapid ENSO transitions perceptually salient.

ENSO variability spans multiple timescales, from intraseasonal oscillations to decadal modulation. To capture this hierarchy, we computed moving averages at window sizes $\omega \in \Omega = \{3, 12, 24, 36\}$ months:
\begin{equation}
\bar{\theta}_\omega(t) = \frac{1}{\omega} \sum_{k=0}^{\omega-1} \theta(t-k),
\label{eq:rolling_mean}
\end{equation}
equivalent to low-pass filtering with frequency response $H(\omega; f) = \sin(\pi f \omega) / [\omega \sin(\pi f)]$ \cite{oppenheim1999discrete}. Each scale was assigned to a separate MIDI track, creating polyphonic textures analogous to gamelan's stratified instrumental layers where different instruments elaborate the same melodic skeleton at different temporal densities \cite{tenzer2000gamelan,Perlman2004}.

We generated eight sonification variants $\mathcal{S}_{i,j}$ indexed by scale type $i \in \{\text{pelog}, \text{slendro}\}$ and composition mode $j$:

\begin{itemize}
    \item \textit{Layered}: all four temporal scales play simultaneously with note durations proportional to window size ($\delta_\omega = \alpha \cdot \Delta t \cdot \omega/12$, where $\alpha = 1.2$ for \textit{pelog}, $1.0$ for \textit{slendro}). This creates dense polyphonic textures with temporal overlap.
    \item \textit{Alternating}: a round-robin scheme activates one temporal scale per time step ($\omega_k = \Omega[t \bmod 4]$), presenting each scale in isolation and introducing forced periodicity in the acoustic output.
    \item \textit{Melodic}: the 12-month average serves as primary voice with the 24-month average as accompaniment at reduced velocity, mimicking gamelan \textit{balungan} (skeletal melody) elaboration \cite{tenzer2000gamelan}.
    \item \textit{Spectral}: Welch periodogram analysis \cite{welch1967use} of the full time series identifies dominant spectral peaks using a Hann window \cite{harris1978use} with 50\% overlap. The relative amplitudes of the top five peaks, normalized to unit sum, weight separate harmonic tracks---encoding ENSO's frequency-domain structure into timbre.
\end{itemize}

All MIDI sequences used tempo $T = 120$ BPM and resolution $\Delta t_{\text{MIDI}} = 0.5$ beats/month, yielding total duration $D = N \cdot \Delta t_{\text{MIDI}} / T \times 60 \approx 7.75$ minutes. MIDI files were rendered to WAV at $f_s = 44.1$ kHz using FluidSynth \cite{fluidsynth} with the FluidR3\_GM SoundFont. The MIDIUtil library \cite{midiutil} handled MIDI generation.

We extracted time-series of perceptually relevant features from rendered WAV files using Librosa \cite{mcfee2015librosa} with hop length $H = 1024$ and frame length $L = 2048$ samples at $f_s = 22{,}050$ Hz (resolutions 46.4 ms and 92.9 ms). These parameters balance the time-frequency uncertainty principle \cite{cohen1995time}.

The spectral centroid (perceptual brightness) at frame $m$ is defined as the frequency-domain center of mass \cite{grey1977multidimensional}:
\begin{equation}
\mathcal{C}[m] = \frac{\sum_{k=0}^{L/2} f[k] \cdot |X[k, m]|}{\sum_{k=0}^{L/2} |X[k, m]|},
\label{eq:spectral_centroid}
\end{equation}
where $X[k,m]$ is the Short-Time Fourier Transform (STFT), $f[k] = k f_s / L$, and $|\cdot|$ the magnitude operator. The RMS energy (perceptual intensity) is \cite{lerch2012introduction}:
\begin{equation}
\mathcal{E}[m] = \sqrt{\frac{1}{L} \sum_{n=0}^{L-1} x^2[mH + n]}.
\label{eq:rms_energy}
\end{equation}

Additional descriptors included zero-crossing rate (ZCR) for high-frequency content \cite{gouyon2000use}, spectral rolloff at $\alpha = 85\%$ \cite{scheirer1997construction}, spectral bandwidth \cite{peeters2004large}, and mel-scaled spectral flux for timbral variability \cite{logan2000mel,tzanetakis2002musical}. For cross-mode comparison, we applied global normalization using pooled 1st/99th percentiles:
\begin{equation}
\mathcal{C}_{\text{norm}}[m] = \frac{\mathcal{C}[m] - \mathcal{C}_{\min}}{\mathcal{C}_{\max} - \mathcal{C}_{\min}}, \quad \mathcal{E}_{\text{norm}}[m] = \frac{\mathcal{E}[m]}{\mathcal{E}_{\max}},
\label{eq:feature_normalization}
\end{equation}
clipped to $[0,1]$. Temporal smoothing with a 20-frame moving average reduced high-frequency noise \cite{mckinney2010data}. Summary statistics (mean, standard deviation, coefficient of variation, skewness, kurtosis), linear trend analysis, and lag-1 autocorrelation \cite{box2015time,seber2012linear} characterized each feature trajectory.

The core analytical framework treats each sonification as a dynamical trajectory in a two-dimensional acoustic state space. We define the state vector
\begin{equation}
\boldsymbol{\Gamma}(m) = \begin{bmatrix} \tilde{\mathcal{C}}_{\text{norm}}(m) \\ \tilde{\mathcal{E}}_{\text{norm}}(m) \end{bmatrix} \in [0,1]^2,
\label{eq:phase_space_state}
\end{equation}
where tildes denote smoothed, normalized features. This choice of coordinates has a natural interpretation: spectral centroid (brightness) correlates with the ``temperature'' of the spectral content, while RMS energy captures the ``intensity'' of the signal. The trajectory $\{\boldsymbol{\Gamma}(m)\}_{m=1}^{M}$ traces how the sonification evolves through coupled brightness-energy dynamics.

We draw on nonlinear dynamics and recurrence quantification analysis (RQA) \cite{eckmann1987recurrence,zbilut1992embeddings,marwan2007recurrence} to characterize these trajectories through the following metrics.

The convex hull area $\mathcal{A} = \text{Area}(\mathcal{H})$, computed via the QuickHull algorithm \cite{barber1996quickhull,preparata1985computational}, quantifies the region of brightness-energy space explored. Larger $\mathcal{A}$ indicates diverse acoustic states; smaller values indicate confinement to restricted parameter regions \cite{trulla1996recurrence}.

The total Euclidean distance traversed,
\begin{equation}
\mathcal{L} = \sum_{m=1}^{M-1} \|\boldsymbol{\Gamma}(m+1) - \boldsymbol{\Gamma}(m)\|_2,
\label{eq:path_length}
\end{equation}
measures cumulative variation in coupled brightness-energy characteristics \cite{webber2015recurrence}.

The ratio $\eta = \mathcal{A} / \mathcal{L}$ captures how efficiently the trajectory covers its accessible phase space. High $\eta$ indicates direct transitions between acoustic states; low values indicate meandering through redundant regions \cite{marwan2007recurrence}.

The temporal mean position $\bar{\boldsymbol{\Gamma}} = M^{-1}\sum_{m}\boldsymbol{\Gamma}(m)$ localizes the average acoustic state; the component-wise standard deviations $\sigma_{\mathcal{C}}$ and $\sigma_{\mathcal{E}}$ quantify dispersion around this centroid.

The Pearson correlation $\rho_{\mathcal{CE}}$ between brightness and energy dimensions \cite{pearson1895note} reveals whether these acoustic axes vary in phase ($\rho > 0$), anti-phase ($\rho < 0$), or independently ($\rho \approx 0$). The sign of this coupling constitutes a dynamical fingerprint of the interaction between scale structure and mapping algorithm.

To quantify quasi-periodicity, we compute the revisit rate---the fraction of trajectory point pairs returning within threshold distance $\epsilon = 0.05$:
\begin{equation}
\mathcal{R}_\epsilon = \frac{2}{N_{\text{pairs}}} \sum_{i=1}^{M-\tau_{\min}} \sum_{j=i+\tau_{\min}}^{M} \mathbb{1}(\|\boldsymbol{\Gamma}(i) - \boldsymbol{\Gamma}(j)\|_2 < \epsilon),
\label{eq:revisit_rate}
\end{equation}
where $\tau_{\min} = 5$ frames excludes trivial consecutive recurrences and $N_{\text{pairs}}$ normalizes appropriately. This metric is a simplified form of the recurrence rate in RQA \cite{zbilut1992embeddings,marwan2007recurrence}, with higher values indicating periodic or quasi-periodic acoustic dynamics analogous to the quasi-periodicity of the underlying ENSO signal. The threshold was chosen to balance sensitivity against noise robustness \cite{mindlin1992topological}. Systematic subsampling (every 10th frame) reduced computational cost from $\mathcal{O}(M^2)$ to $\mathcal{O}((M/10)^2)$ while preserving temporal structure given the 20-frame smoothing window.

All trajectory metrics were computed using NumPy \cite{harris2020array} and SciPy \cite{virtanen2020scipy}. For each variant $\mathcal{S}_{i,j}$, the metric set $\{\mathcal{A}, \mathcal{L}, \eta, \bar{\boldsymbol{\Gamma}}, \boldsymbol{\sigma}_{\Gamma}, \rho_{\mathcal{CE}}, \mathcal{R}_\epsilon\}$ enables comparison across scale families and composition strategies.

\section{Results}

The Ni\~{n}o 3.4 index exhibited mean $\bar{\theta} = -0.098$\textdegree C, standard deviation $\sigma = 0.775$\textdegree C, and range 5.06\textdegree C (maximum 2.57\textdegree C in November 2015; minimum $-2.49$\textdegree C in January 1890). The distribution showed positive skewness $\gamma_1 = 0.474$ and excess kurtosis $\gamma_2 = 0.382$, confirming the well-established warm-phase amplitude asymmetry arising from ENSO's nonlinear dynamics \cite{an2005nonlinearity} and heavier-than-Gaussian tails. Lag-1 autocorrelation was $\rho_1 = 0.926$, reflecting strong thermal inertia.

\begin{figure}[H]
\centering
\includegraphics[width=\textwidth]{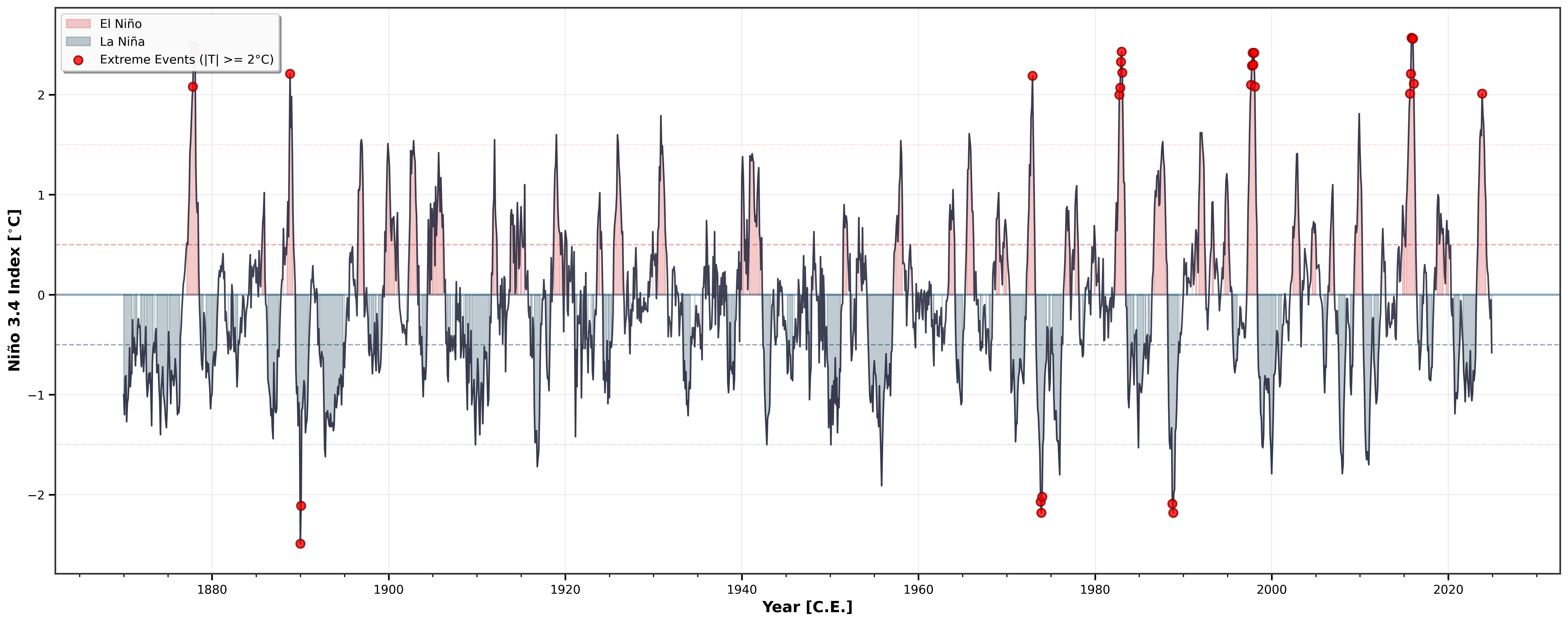}
\caption{Ni\~{n}o 3.4 SST anomaly index, January 1870 to December 2024 ($N = 1{,}860$). Red/blue shading: El Ni\~{n}o/La Ni\~{n}a conditions ($\pm 0.5$\textdegree C thresholds). Red circles: extreme events ($|\theta| \geq 2.0$\textdegree C). The time series exhibits the irregular quasi-periodicity, amplitude asymmetry, and multi-decadal modulation characteristic of low-dimensional climate chaos.}
\label{fig:enso_timeseries}
\end{figure}

ENSO phase partitioning identified El Ni\~{n}o conditions during 377 months (20.3\%, mean 1.05\textdegree C), La Ni\~{n}a during 597 months (32.1\%, mean $-0.917$\textdegree C), and neutral conditions during 886 months (47.6\%). The asymmetry in duration---43 El Ni\~{n}o episodes (mean 8.7 months) versus 57 La Ni\~{n}a episodes (mean 11.4 months)---reflects the nonlinear recharge-discharge dynamics where the discharge phase (La Ni\~{n}a) relaxes more slowly than the growth phase (El Ni\~{n}o). The longest La Ni\~{n}a episodes reached 33 months, while the longest El Ni\~{n}o lasted 17 months. Month-to-month rate of change statistics showed $\sigma_{\Delta\theta} = 0.299$\textdegree C/month with extremes of $+1.13$\textdegree C/month (December 1939) and $-1.24$\textdegree C/month (December 1889). The complete time series (Figure~\ref{fig:enso_timeseries}) displays the irregular oscillatory behavior, amplitude modulation, and intermittent extreme events characteristic of ENSO as a complex dynamical system.

Acoustic features extracted from the eight sonification variants showed systematic differentiation across both scale families and composition modes (Table~\ref{tab:music_char}). Mean spectral centroid ranged from 809 Hz (\textit{pelog} melodic) to 1,863 Hz (\textit{slendro} layered). The \textit{slendro} family produced consistently higher brightness values across all composition modes---a consequence of its more uniform interval spacing distributing pitch content across a wider spectral range. Mean RMS energy varied from 0.008 (\textit{slendro} melodic) to 0.022 (\textit{slendro} layered), with layered modes exhibiting highest intensities regardless of scale family, consistent with their polyphonic texture. Dynamic range spanned 2.08 to 3.63, with \textit{slendro} melodic showing the widest contrast.

\begin{table}[H]
\centering
\caption{Acoustic feature statistics for eight sonification variants. Brightness: mean spectral centroid (Hz); Intensity: mean RMS energy; DynRng: dynamic range ratio; Percuss: mean zero-crossing rate; Flux: mean spectral flux.}
\label{tab:music_char}
\small
\begin{tabular}{lrrrrr}
\hline
Mode & Brightness [Hz] & Intensity & DynRng & Percuss & Flux \\
\hline
\textit{Pelog} Layered & 1,238 & 0.021 & 2.26 & 0.065 & 664 \\
\textit{Pelog} Alternating & 1,064 & 0.013 & 2.94 & 0.055 & 489 \\
\textit{Pelog} Melodic & 809 & 0.018 & 2.18 & 0.041 & 568 \\
\textit{Pelog} Spectral & 1,270 & 0.012 & 2.96 & 0.053 & 988 \\
\textit{Slendro} Layered & 1,863 & 0.022 & 2.08 & 0.134 & 985 \\
\textit{Slendro} Alternating & 1,233 & 0.011 & 3.48 & 0.074 & 757 \\
\textit{Slendro} Melodic & 1,418 & 0.008 & 3.63 & 0.055 & 866 \\
\textit{Slendro} Spectral & 1,274 & 0.012 & 2.75 & 0.054 & 986 \\
\hline
\end{tabular}
\end{table}

\begin{figure}[H]
\centering
\includegraphics[width=\textwidth]{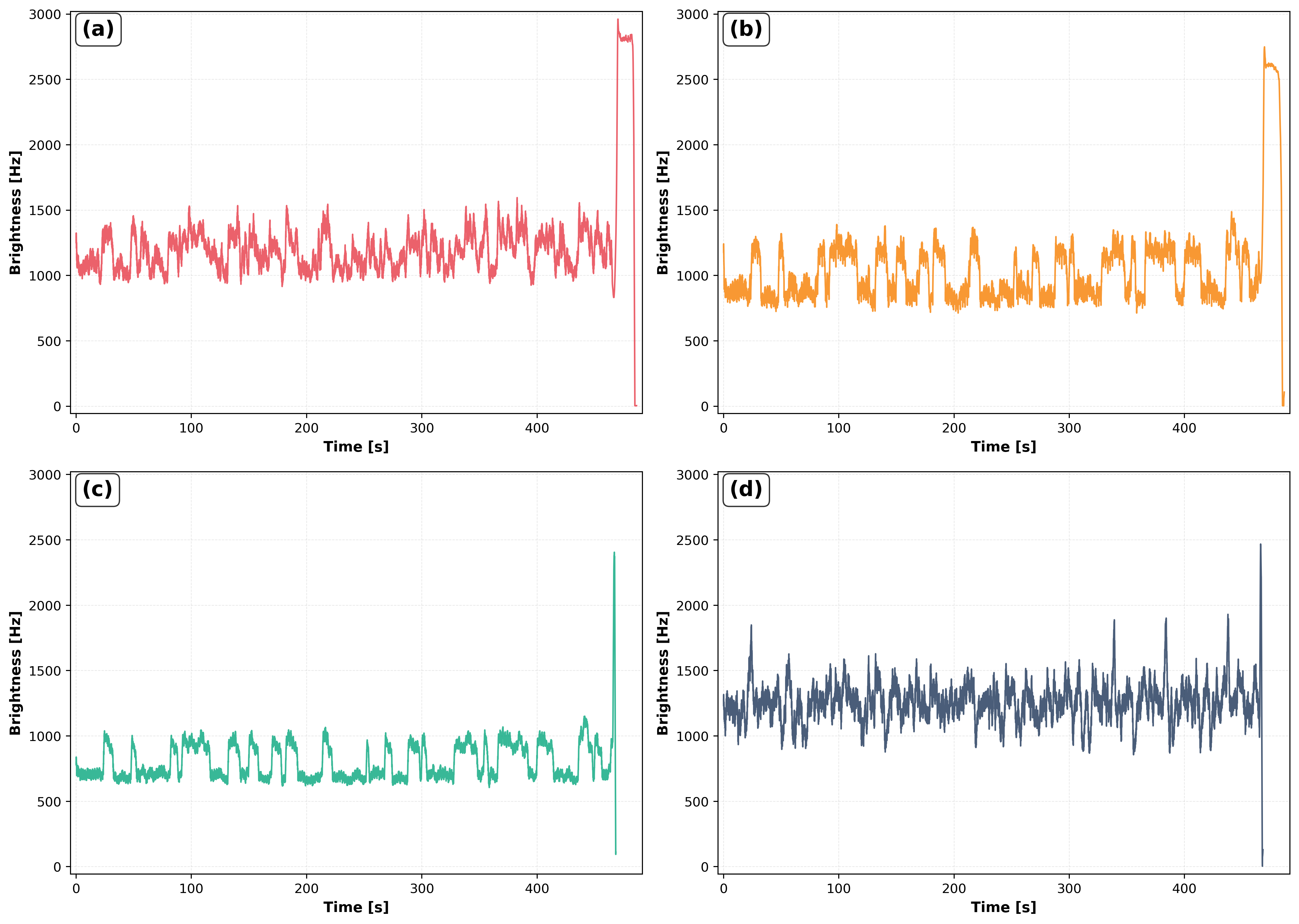}
\caption{Spectral centroid trajectories for \textit{pelog} modes: (a) layered, (b) alternating, (c) melodic, (d) spectral. Terminal descents near $t \approx 470$ s are fade-out artifacts.}
\label{fig:brightness_pelog}
\end{figure}

\begin{figure}[H]
\centering
\includegraphics[width=\textwidth]{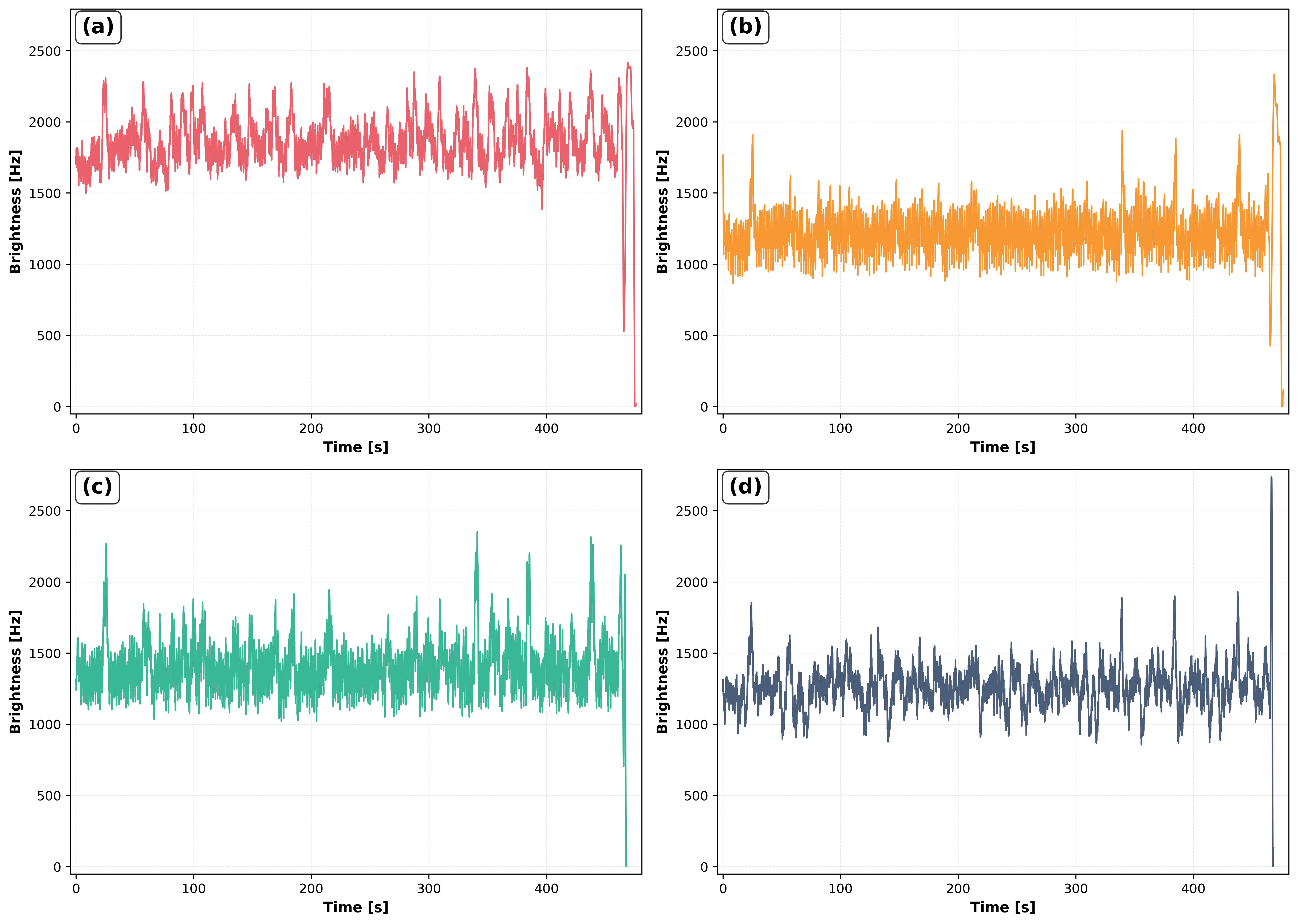}
\caption{Spectral centroid trajectories for \textit{slendro} modes. \textit{Slendro} layered (a) achieves the highest mean brightness (1,863 Hz) of all variants.}
\label{fig:brightness_slendro}
\end{figure}

The temporal evolution of spectral centroid trajectories (Figures~\ref{fig:brightness_pelog} and \ref{fig:brightness_slendro}) revealed mode-specific dynamics over the $\sim$470--490 second composition durations. Key patterns include: layered modes maintained relatively stable brightness with high-frequency fluctuations; alternating modes showed sustained oscillations reflecting the forced round-robin periodicity; melodic modes exhibited the most constrained ranges; and spectral modes displayed complex multi-scale variation. The \textit{slendro} layered mode occupied consistently higher frequency space than any \textit{pelog} variant.

\begin{figure}[H]
\centering
\includegraphics[width=\textwidth]{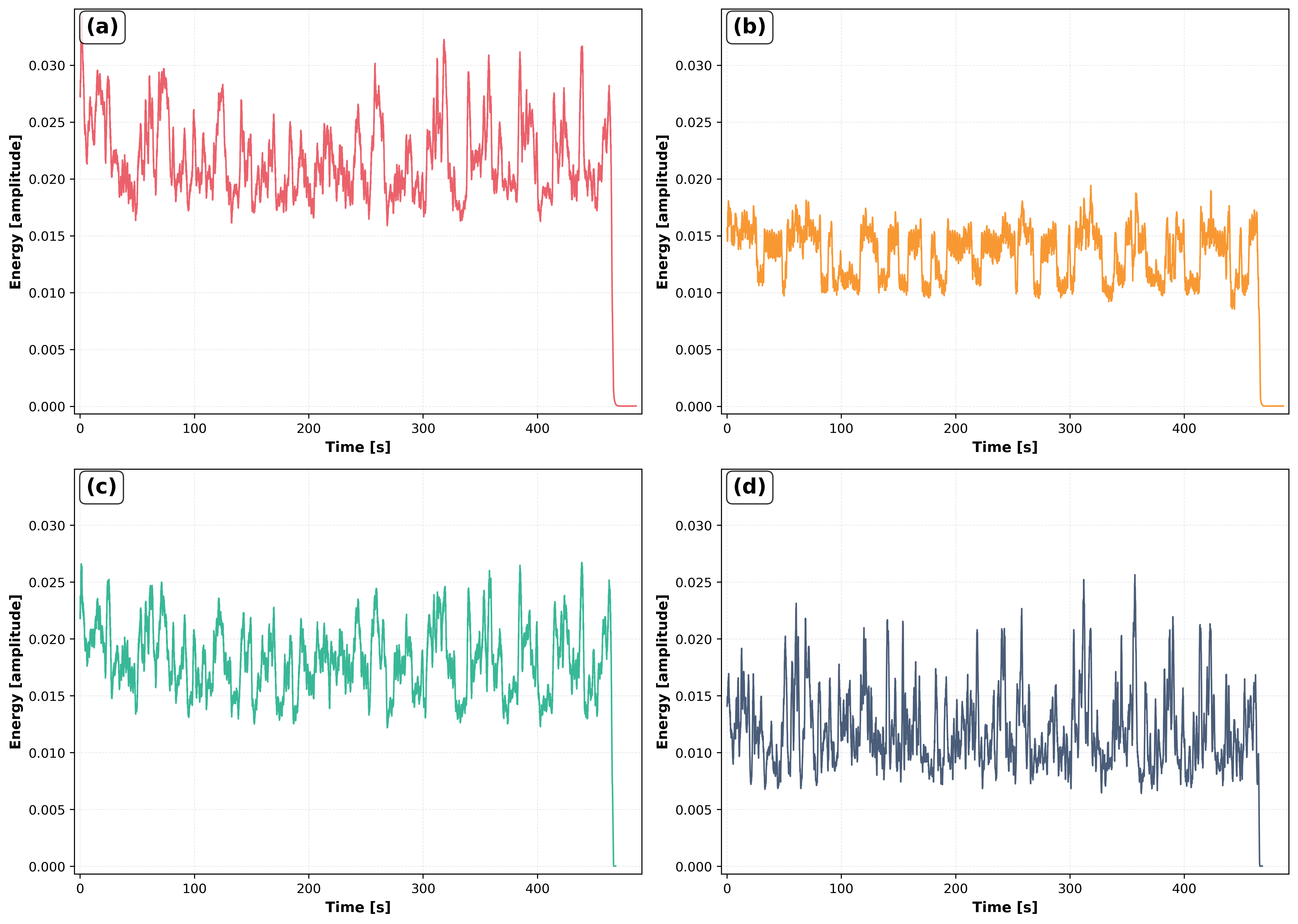}
\caption{RMS energy trajectories for \textit{pelog} modes. All modes show significant negative trends ($p < 0.001$).}
\label{fig:energy_pelog}
\end{figure}

\begin{figure}[H]
\centering
\includegraphics[width=\textwidth]{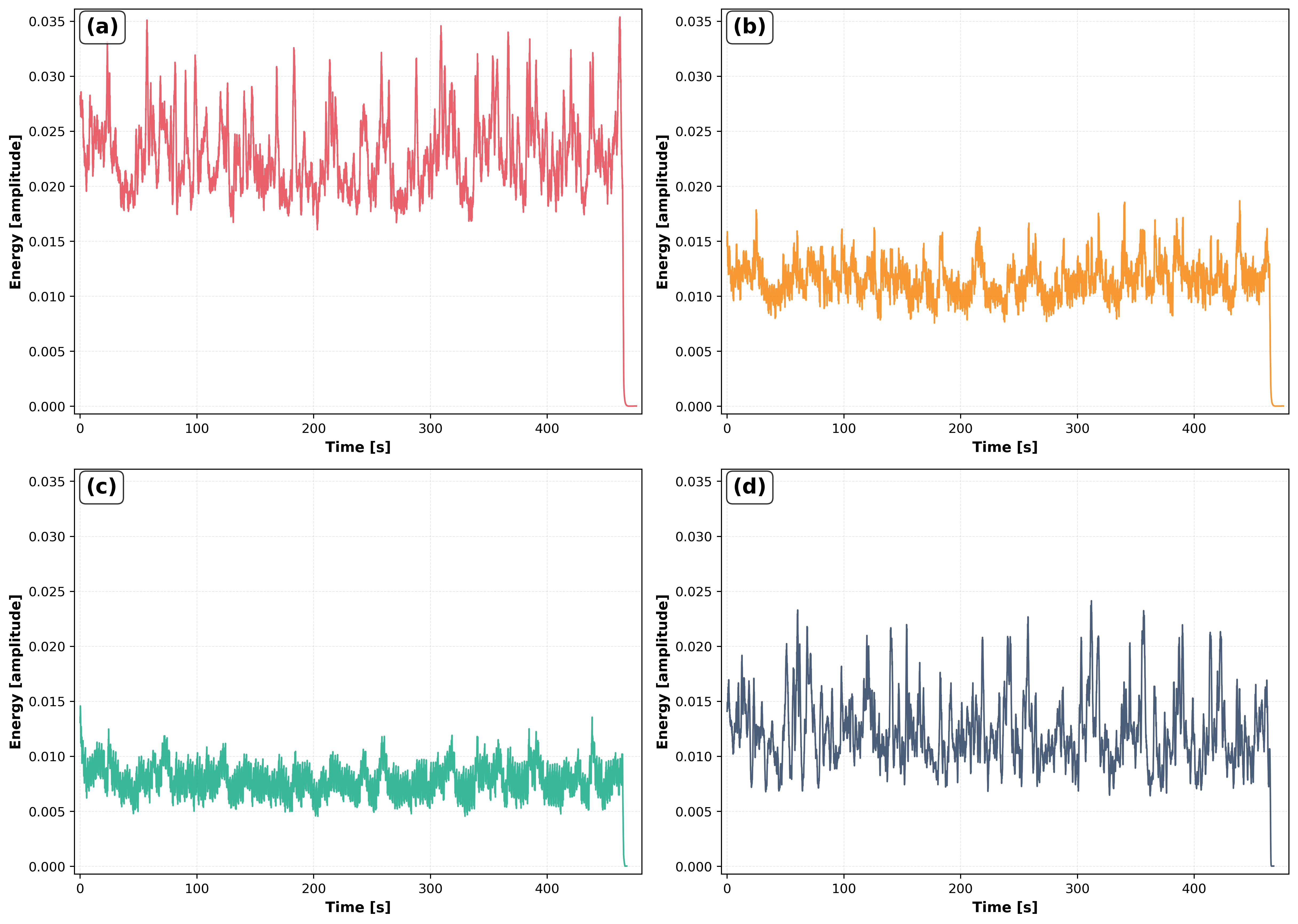}
\caption{RMS energy trajectories for \textit{slendro} modes. \textit{Slendro} layered (a) achieves the highest mean energy (0.022) across all variants.}
\label{fig:energy_slendro}
\end{figure}

RMS energy trajectories (Figures~\ref{fig:energy_pelog} and \ref{fig:energy_slendro}) displayed complementary temporal patterns. Layered modes maintained the highest amplitudes; melodic modes the lowest. All modes exhibited statistically significant negative energy trends ($p < 0.001$), likely an artifact of the temporal compression mapping.

Statistical characterization (Table~\ref{tab:timeseries_stats}) confirmed systematic scale-family differences: \textit{pelog} modes averaged 1,095 Hz brightness ($\sigma = 234$ Hz, CV = 0.215), while \textit{slendro} averaged 1,447 Hz ($\sigma = 188$ Hz, CV = 0.132). The lower coefficient of variation in \textit{slendro} indicates more stable relative brightness despite higher absolute values. All modes showed lag-1 autocorrelation exceeding 0.958 for brightness and 0.981 for energy, indicating highly persistent feature evolution that parallels the strong persistence ($\rho_1 = 0.926$) of the input ENSO signal. This persistence survival through the sonification pipeline suggests the mapping partially preserves the temporal correlation structure of the source dynamics.

\begin{table}[H]
\centering
\caption{Time series statistics for spectral centroid and RMS energy (representative subset). $\bar{x}$: mean; $\sigma_x$: standard deviation; $\gamma_x$: coefficient of variation; $\beta$: trend slope per frame; $R^2$: determination coefficient; $\rho_1$: lag-1 autocorrelation.}
\label{tab:timeseries_stats}
\small
\begin{tabular}{llrrrrrr}
\hline
Variable & Mode & $\bar{x}$ & $\sigma_x$ & $\gamma_x$ & $\beta$ & $R^2$ & $\rho_1$ \\
\hline
\multirow{4}{*}{\shortstack[l]{Brightness\\{[Hz]}}}
& \textit{Pelog} Layered & 1,238 & 311 & 0.251 & 0.031 & 0.091 & 0.993 \\
& \textit{Pelog} Spectral & 1,270 & 157 & 0.123 & 0.004 & 0.005 & 0.958 \\
& \textit{Slendro} Layered & 1,863 & 208 & 0.112 & 0.002 & 0.001 & 0.980 \\
& \textit{Slendro} Spectral & 1,274 & 162 & 0.127 & 0.004 & 0.006 & 0.959 \\
\hline
\multirow{4}{*}{\shortstack[l]{Energy\\{[-]}}}
& \textit{Pelog} Layered & 0.021 & 0.005 & 0.258 & $-6.1\times10^{-7}$ & 0.118 & 0.998 \\
& \textit{Pelog} Spectral & 0.012 & 0.003 & 0.284 & $-1.1\times10^{-7}$ & 0.008 & 0.996 \\
& \textit{Slendro} Layered & 0.022 & 0.005 & 0.213 & $-2.6\times10^{-7}$ & 0.026 & 0.996 \\
& \textit{Slendro} Spectral & 0.012 & 0.003 & 0.266 & $-1.2\times10^{-7}$ & 0.011 & 0.996 \\
\hline
\end{tabular}
\end{table}

\begin{figure}[H]
\centering
\includegraphics[width=\textwidth]{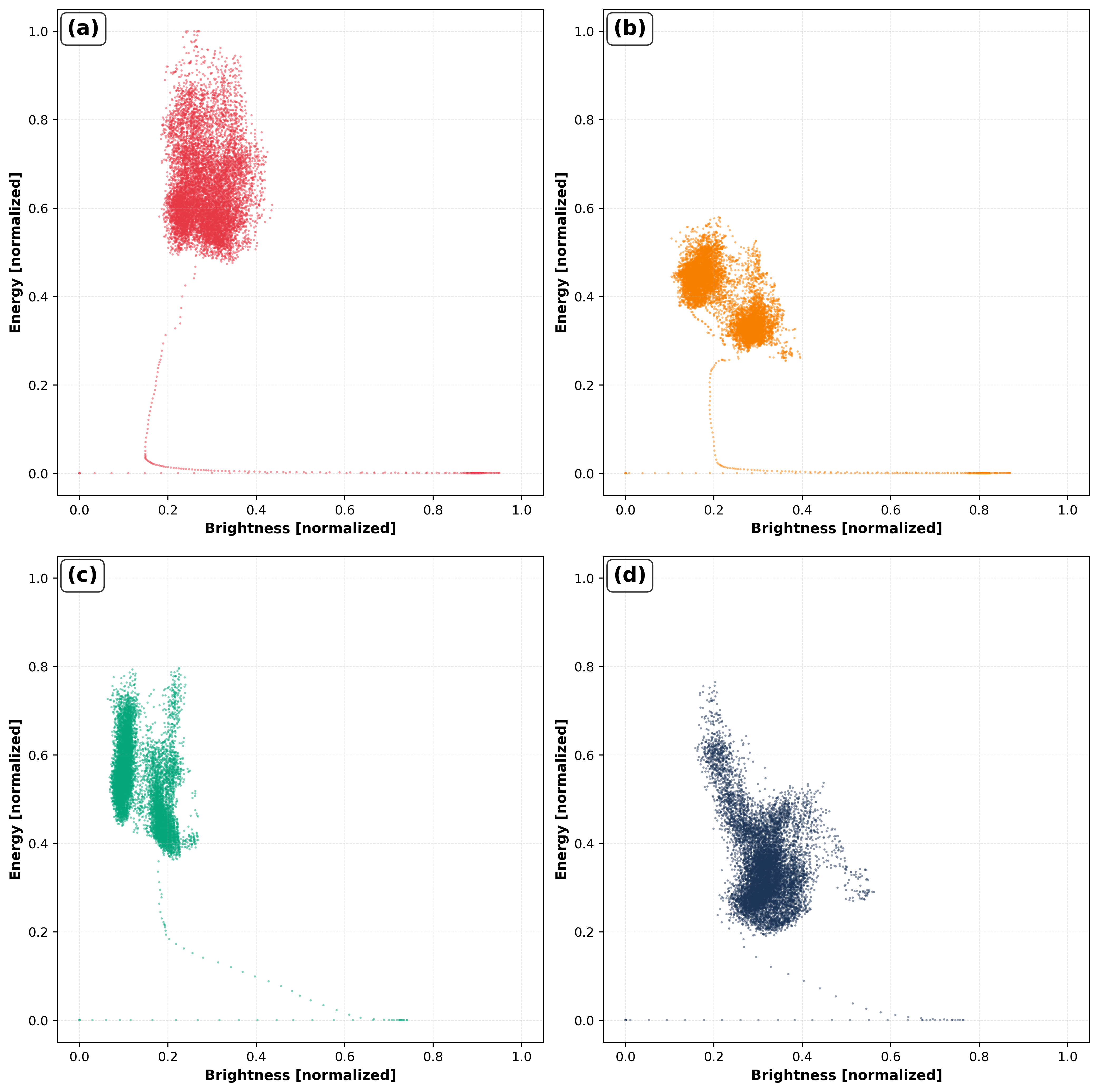}
\caption{Phase space trajectories for \textit{pelog} variants. Each point represents $\boldsymbol{\Gamma}(m) = [\tilde{\mathcal{C}}_{\text{norm}}(m), \tilde{\mathcal{E}}_{\text{norm}}(m)]$. Colors: red (layered), orange (alternating), green (melodic), navy (spectral).}
\label{fig:phase_pelog}
\end{figure}

\begin{figure}[H]
\centering
\includegraphics[width=\textwidth]{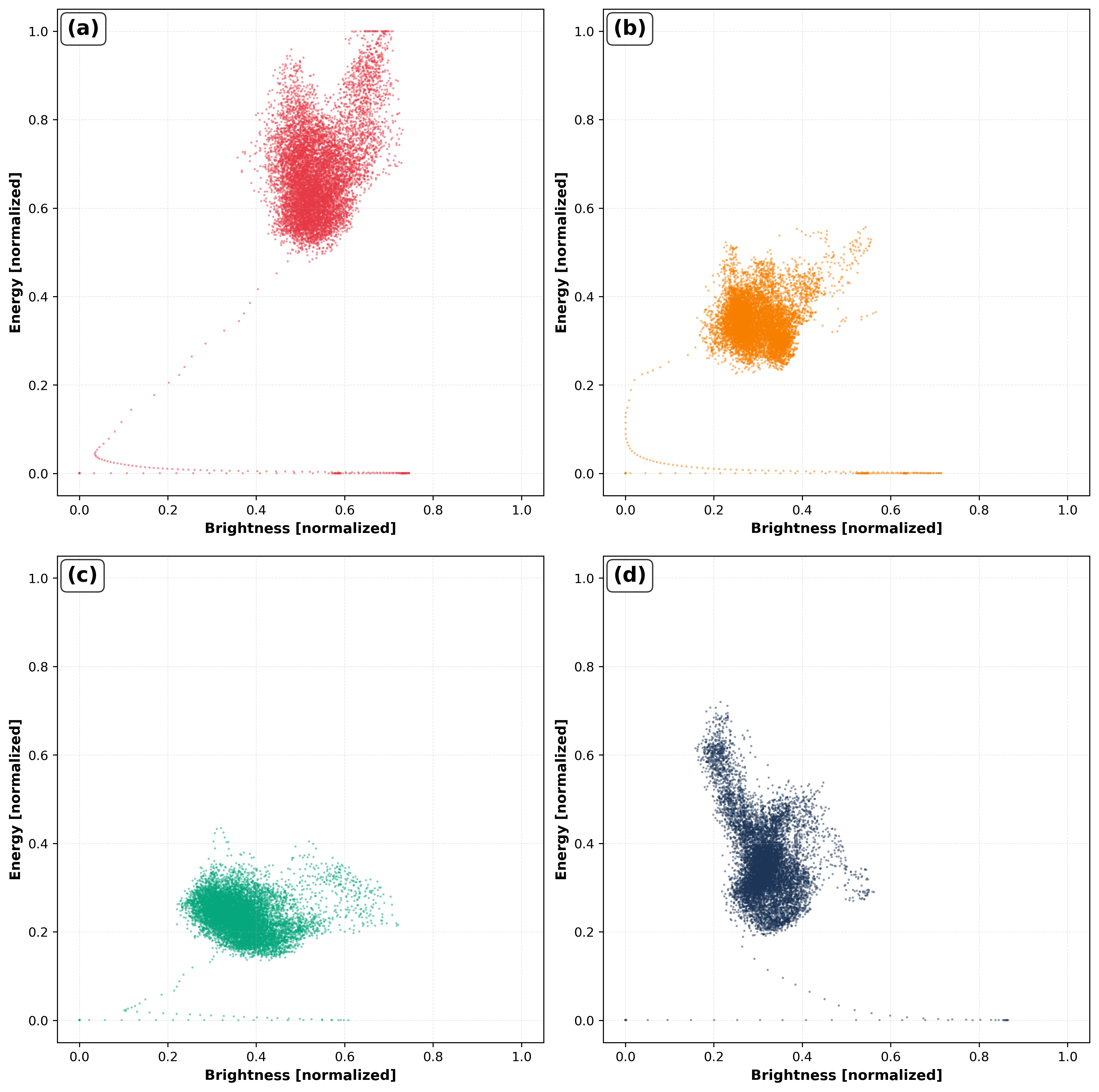}
\caption{Phase space trajectories for \textit{slendro} variants. Note the clear spatial separation of layered mode (a) from other modes---a stronger inter-mode differentiation than observed in the \textit{pelog} family.}
\label{fig:phase_slendro}
\end{figure}

Phase space trajectories in normalized brightness-energy coordinates revealed geometrically distinct structures for each variant (Figures~\ref{fig:phase_pelog} and \ref{fig:phase_slendro}). \textit{Pelog} layered occupied a broad region ($\tilde{\mathcal{C}}_{\text{norm}} \approx 0.3$--$0.7$, $\tilde{\mathcal{E}}_{\text{norm}} \approx 0.5$--$1.0$) with dense concentration at mid-to-high energy. \textit{Pelog} melodic showed the most restricted extent ($\tilde{\mathcal{C}}_{\text{norm}} \approx 0.1$--$0.3$, $\tilde{\mathcal{E}}_{\text{norm}} \approx 0.15$--$0.35$). Spectral modes in both families exhibited characteristic vertical elongation, reflecting energy variability at relatively stable brightness. The \textit{slendro} layered mode occupied a distinct high-brightness, high-energy region ($\tilde{\mathcal{C}}_{\text{norm}} \approx 0.45$--$0.65$, $\tilde{\mathcal{E}}_{\text{norm}} \approx 0.5$--$0.9$), well separated from other modes.

Quantitative trajectory metrics (Table~\ref{tab:phase_metrics}) revealed substantial mode-dependent variation. Convex hull areas ranged from 0.211 (\textit{slendro} melodic) to 0.529 (\textit{pelog} layered). Total path lengths varied from 99.8 (\textit{pelog} alternating) to 183 (\textit{slendro} layered and melodic). The exploration index was uniformly low ($\eta \in [0.001, 0.004]$), indicating that all trajectories are highly meandering---they traverse far more path length than needed to cover their phase space footprint. This is consistent with the source signal's quasi-oscillatory character: the trajectory repeatedly revisits similar acoustic states as the ENSO data cycles between warm and cold phases.

\begin{table}[H]
\centering
\caption{Phase space trajectory metrics. $\mathcal{A}$: convex hull area; $\mathcal{L}$: path length; $\eta$: exploration index ($\mathcal{A}/\mathcal{L}$); $\bar{\mathcal{C}}_n, \bar{\mathcal{E}}_n$: centroid coordinates; $\rho_{\mathcal{CE}}$: brightness-energy correlation; $\mathcal{R}_\epsilon$: revisit rate ($\epsilon = 0.05$). All correlations significant at $p < 0.001$.}
\label{tab:phase_metrics}
\small
\begin{tabular}{lrrrrrrc}
\hline
Mode & $\mathcal{A}$ & $\mathcal{L}$ & $\eta$ & $\bar{\mathcal{C}}_n$ & $\bar{\mathcal{E}}_n$ & $\rho_{\mathcal{CE}}$ & $\mathcal{R}_\epsilon$ \\
\hline
\textit{Pelog} Layered & 0.529 & 147 & 0.004 & 0.302 & 0.621 & $-$0.568 & 0.119 \\
\textit{Pelog} Alternating & 0.285 & 99.8 & 0.003 & 0.236 & 0.387 & $-$0.790 & 0.240 \\
\textit{Pelog} Melodic & 0.356 & 111 & 0.003 & 0.140 & 0.537 & $-$0.513 & 0.183 \\
\textit{Pelog} Spectral & 0.333 & 166 & 0.002 & 0.314 & 0.354 & $-$0.362 & 0.163 \\
\textit{Slendro} Layered & 0.488 & 183 & 0.003 & 0.537 & 0.665 & $+$0.317 & 0.123 \\
\textit{Slendro} Alternating & 0.301 & 140 & 0.002 & 0.300 & 0.338 & $-$0.118 & 0.230 \\
\textit{Slendro} Melodic & 0.211 & 183 & 0.001 & 0.369 & 0.235 & $-$0.089 & 0.225 \\
\textit{Slendro} Spectral & 0.346 & 168 & 0.002 & 0.315 & 0.363 & $-$0.405 & 0.168 \\
\hline
\end{tabular}
\end{table}

Three dynamical signatures deserve emphasis:
\begin{itemize}
    \item \textit{Recurrence structure}: Alternating modes exhibited the highest revisit rates in both scale families ($\mathcal{R}_\epsilon = 0.240$ for \textit{pelog}, 0.230 for \textit{slendro}), while layered modes showed the lowest (0.119, 0.123). The round-robin composition strategy introduces a period-4 forcing in acoustic space, creating a more recurrent trajectory---effectively amplifying the quasi-periodic character of the ENSO input. Layered polyphony, by superposing all temporal scales simultaneously, produces a higher-dimensional acoustic signal that projects into the 2D brightness-energy plane as a less recurrent, more space-filling trajectory. This distinction between low-recurrence/high-area (layered) and high-recurrence/low-area (alternating) trajectories parallels the classic distinction between space-filling and periodic orbits in dynamical systems \cite{strogatz2015nonlinear}.
    \item \textit{Brightness-energy coupling}: The \textit{pelog} family showed predominantly strong negative correlations ($\bar{\rho}_{\mathcal{CE}} = -0.558$, especially $-0.790$ in alternating mode), indicating that brightness and energy vary in anti-phase. The \textit{slendro} family showed weak mean correlation ($\bar{\rho}_{\mathcal{CE}} = -0.074$), with layered mode uniquely exhibiting positive coupling ($+0.317$). This bifurcation in coupling regime between scale families constitutes an emergent property of the interaction between interval structure and the sonification mapping---neither the input data nor the mapping algorithm alone determines the sign of the correlation.
    \item \textit{Scale-family separation}: Aggregating by family, \textit{pelog} modes averaged $\mathcal{A} = 0.376$, $\mathcal{L} = 131$, $\mathcal{R}_\epsilon = 0.176$; \textit{slendro} averaged $\mathcal{A} = 0.337$, $\mathcal{L} = 169$, $\mathcal{R}_\epsilon = 0.187$. The \textit{slendro} family thus produces longer trajectories with marginally higher recurrence rates but covering less phase space area---a more ``wandering but revisiting'' dynamic compared to \textit{pelog}'s ``expansive but less recurrent'' character.
\end{itemize}

\section{Discussion}

Our results demonstrate that sonification of a complex dynamical system through culturally-grounded musical frameworks produces acoustic output whose phase space properties can be rigorously characterized using tools from nonlinear dynamics. This framing addresses an underappreciated aspect of sonification research: the sonification pipeline itself is a complex mapping whose output dynamics depend on the interaction between source signal properties, the mapping function, and the musical framework employed. The phase space analysis reveals that this interaction generates emergent properties---particularly the coupling regime between brightness and energy dimensions---not predictable from any component in isolation.

The phase space metrics we employ---convex hull area, path length, exploration index, recurrence rate, and coupling correlation---form a geometric characterization toolkit applicable to any sonification design, not just gamelan-based approaches. The distinction between high-recurrence/low-area trajectories (alternating modes) and low-recurrence/high-area trajectories (layered modes) offers a principled framework for choosing composition strategies based on desired dynamical properties of the acoustic output. If quasi-periodicity is the target property to communicate, alternating modes are preferable; if diversity of acoustic states is the goal, layered modes are indicated.

A central question for any data representation method is what structural properties of the source survive the transformation. We observe partial preservation: the high lag-1 autocorrelation of the ENSO signal ($\rho_1 = 0.926$) appears as high persistence in acoustic feature trajectories ($\rho_1 > 0.958$ for all modes). The quasi-periodic character of ENSO maps to nonzero revisit rates in the acoustic phase space, though the specific recurrence structure depends heavily on composition mode. The ENSO distribution's positive skewness ($\gamma_1 = 0.474$) is not directly preserved in brightness distributions, which show variable skewness including both positive and negative values---an expected consequence of the nonlinear pitch-mapping function. Full characterization of information transfer through the sonification pipeline would require mutual information estimation between input and output signals, an important direction for future work.

The systematic trends we observe---increasing brightness and decreasing energy over the composition duration---represent an artifact of temporal compression rather than a preserved ENSO property. The 155-year record compressed to 7.75 minutes may induce non-stationary acoustic features through synthesis-engine effects unrelated to the input data. Prior work demonstrates that data density significantly affects auditory comprehension \cite{Nees2008}, suggesting this compression ratio warrants systematic optimization.

The striking bifurcation in brightness-energy coupling between scale families---strong anti-phase in \textit{pelog} versus near-independence in \textit{slendro}---is an emergent property that arises from the interaction between the interval structure and the parameter-mapping algorithm. The \textit{pelog} intervals $\{0, 1, 3, 7, 8\}$ produce clustered pitch distributions where moving to higher scale degrees (brighter spectral content) concentrates spectral energy differently than the more distributed \textit{slendro} intervals $\{0, 2, 3, 7, 9\}$. This finding has implications beyond our specific application: it demonstrates that the choice of musical scale is not merely aesthetic but introduces structural constraints on the dynamical properties of sonified output.

Music itself has long been understood as exhibiting complex systems properties. Voss and Clarke's foundational observation that pitch and loudness fluctuations in music display $1/f$ spectral scaling \cite{voss1975} established that musical signals occupy a regime between uncorrelated randomness and perfectly periodic order---exactly the regime characteristic of many natural complex systems including ENSO. Our gamelan-mapped ENSO sonifications inherit complexity from both the source signal and the musical framework, and the phase space analysis provides tools for disentangling these contributions.

The choice of gamelan is not incidental to the complex systems framing. Gamelan's stratified polyphony---where instruments elaborate a common melodic skeleton at different temporal densities \cite{tenzer2000gamelan,Perlman2004}---is itself a multi-scale hierarchical system. The nested cyclical time organization \cite{Becker1981} produces temporal structures fundamentally different from the linear-progressive organization of Western music. Our multi-scale decomposition ($\omega \in \{3, 12, 24, 36\}$ months) maps naturally onto this hierarchical structure, with shorter windows capturing fast fluctuations and longer windows isolating slower dynamics---a temporal separation analogous to how different gamelan instruments relate to the colotomic cycle.

However, our implementation involves significant simplifications. The 12-tone equal temperament approximation discards the microtonal \textit{embat} that makes each gamelan set unique \cite{surjodiningrat1972tone}. General MIDI synthesis lacks the timbral characteristics of bronze metallophone instruments achievable through specialized techniques \cite{Schiemer2016}. The \textit{pathet} modal system, which governs temporal progression and carries affective meaning in traditional practice \cite{Perlman1998}, is not incorporated. Most critically, we have not engaged with Indonesian communities to assess cultural resonance---effective communication requires design conscious of audience relationship to content \cite{Lenzi2020}.

Several limitations constrain this work. The phase space analysis characterizes the sonification output, not the fidelity of the data-to-sound mapping. Metrics like mutual information or transfer entropy between input and output would strengthen the dynamical systems analysis. Full RQA---including determinism, laminarity, and longest diagonal line measures \cite{zbilut1992embeddings,marwan2007recurrence}---would provide richer dynamical characterization than our simplified recurrence rate. The threshold $\epsilon = 0.05$ was fixed rather than optimized systematically. The 10-level pitch quantization ($\sim$0.6\textdegree C resolution) is coarse, potentially discarding meaningful climate variability.

Most critically, we lack perceptual validation. Quantitative acoustic characterization cannot substitute for listener studies assessing pattern recognition, event identification, or data comprehension \cite{Kramer1999,Hermann2013}. Musical complexity may hinder rather than help engagement \cite{Middleton2023}. We contribute the analytical framework for evaluating sonification designs; the communicative effectiveness question remains open.

Future work should address these gaps through information-theoretic measures of mapping fidelity, full RQA with systematic embedding parameter selection, perceptual experiments comparing composition modes, and extension to multivariate ENSO representations mapping SST patterns, atmospheric pressure, and thermocline depth to gamelan's multi-layered instrumental architecture.

\section{Conclusion}

We demonstrate that traditional Javanese gamelan scales provide a viable framework for ENSO data sonification, producing eight acoustically distinct variants whose phase space dynamics can be rigorously characterized using tools from nonlinear dynamics. The key findings are: (1) \textit{slendro} modes generate consistently higher spectral centroids than \textit{pelog}, reflecting the influence of interval structure on spectral properties; (2) layered compositions explore broader phase space with lower recurrence, while alternating modes exhibit the highest trajectory revisit rates, paralleling the distinction between space-filling and periodic orbits; (3) the two scale families induce qualitatively different coupling regimes between brightness and energy dimensions, an emergent property of the scale-mapping interaction; and (4) multi-scale temporal decomposition encodes ENSO variability across subseasonal to decadal timescales, leveraging gamelan's inherent hierarchical temporal structure. These results establish that phase space trajectory analysis provides geometric tools for systematic comparison of sonification designs within a complex systems framework, and that culturally-grounded musical systems can serve as structured dynamical mappings between climate variability and auditory perception.

\section*{Acknowledgments}
This work was funded by ITB 3P Research Program 2026 (Top Tier Scheme) through the Directorate of Research and Innovation, Bandung Institute of Technology (Project ID: FITB.PN-6--198--2026).

\section*{Author Contributions}

S.H.S.H.: Conceptualization; Data curation; Formal analysis; Investigation; Methodology; Software; Validation; Visualization; Writing -- original draft; Writing -- review \& editing. R.S.: Conceptualization; Formal analysis; Investigation; Funding acquisition; Project administration; Supervision; Writing -- review \& editing. N.J.T.: Supervision; Writing -- review \& editing. I.P.A.: Supervision; Writing -- review \& editing. F.R.F.: Supervision; Writing -- review \& editing. All authors have read and approved the final manuscript.

\section*{Open Research}

The Ni\~{n}o 3.4 SST anomaly index (January 1870--December 2024) from HadISST 1.1 is available at \url{https://psl.noaa.gov/data/timeseries/month/DS/Nino34/}. Python scripts and processed datasets are archived at \url{https://github.com/sandyherho/suppl-enso-javanese-sonification} under WTFPL. All eight sonification variants (MIDI and WAV) are deposited at \url{https://doi.org/10.17605/OSF.IO/QY82M} under AFL 3.0.

\end{document}